\documentclass[11pt]{article}

\usepackage{amsmath}
\usepackage{graphicx}
\usepackage{amsfonts}
\usepackage{amssymb}
\usepackage{epsfig}
\usepackage{color}
\usepackage{psfrag}

\newcommand{\EQ}{\begin{equation}}
\newcommand{\EN}{\end{equation}}
\newcommand{\be}{\begin{equation}}
\newcommand{\ee}{\end{equation}}
\newcommand{\bea}{\begin{eqnarray}}
\newcommand{\eea}{\end{eqnarray}}

\newcommand \lab {\label}
\newcommand{\var}{\varepsilon}

\newcommand{\goto}{\rightarrow}

\def\goto{\longrightarrow}
                                                                                                                               
\begin{document} \setcounter{page}{0} 
\topmargin 0pt
\oddsidemargin 5mm
\renewcommand{\thefootnote}{\arabic{footnote}}
\newpage
\setcounter{page}{0}
\topmargin 0pt
\oddsidemargin 5mm
\renewcommand{\thefootnote}{\arabic{footnote}}
\newpage
\begin{titlepage}
\begin{flushright}
SISSA 53/2005/FM \\
PM/05-21
\end{flushright}
\vspace{0.5cm}
\begin{center}
{\large {\bf Decay of particles above threshold in the}}\\ 
{\large {\bf Ising field theory with magnetic field}}\\
\vspace{1.8cm}
{\large Gesualdo Delfino$^{a,b}$, Paolo Grinza$^c$ and 
Giuseppe Mussardo$^{a,b}$}\\
\vspace{0.5cm}
{\em ${}^a\,$International School for Advanced Studies (SISSA),}\\ 
{\em via Beirut 2-4, 34014 Trieste, Italy}\\
{\em ${}^b\,$Istituto Nazionale di Fisica Nucleare, sezione di Trieste, Italy}\\
\vspace{1mm}
{\em ${}^c\,$Laboratoire de Physique Th\'eorique et Astroparticules, Universit\'e 
Montpellier II,}\\
{\em Place Eug\`ene Bataillon, 34095 Montpellier Cedex 05, France}\\ 
\end{center}
\vspace{1.2cm}

\renewcommand{\thefootnote}{\arabic{footnote}}
\setcounter{footnote}{0}

\begin{abstract}
\noindent
The two-dimensional scaling Ising model in a magnetic field at critical 
temperature is integrable and possesses eight stable particles $A_i$ ($i=1,
\ldots,8$) with different masses. The heaviest five lie above threshold and owe
their stability to integrability. We use form factor perturbation theory to 
compute the decay widths of the first two particles above threshold when 
integrability is broken by a small deviation from the critical temperature. The
lifetime ratio $t_4/t_5$ is found to be 0.233; the particle $A_5$ decays at 
47\% in the channel $A_1A_1$ and for the remaining fraction in the channel 
$A_1A_2$. The increase of the lifetime with the mass, a feature which can be 
expected in two dimensions from phase space considerations, is in 
this model further enhanced by the dynamics.

\end{abstract}
\end{titlepage}

\newpage

\section{Spectrum and analyticity in Ising field theory}
Quantum field theory in $1+1$ dimensions allows for a subclass of integrable
cases. While integrability is hardly detectable at the level of correlation
functions, it has very visible implications at the scattering level. Indeed,
the presence of infinitely many conserved quantities forces the $S$-matrix to
be completely elastic and factorized \cite{ZZ}. These properties, in turn, 
allow for a feature of the particle spectrum which is peculiar to integrable 
quantum field theories: the existence of particles with mass higher than the 
lowest threshold which are stable even in the absence of an internal symmetry 
which would prevent them from decaying. Once integrability is broken by an 
arbitrarily small perturbation, the stability of these particles would 
contradict though the analyticity requirements on the $S$-matrix, so that they 
are forced to develop a finite lifetime. This phenomenon already appears in the
field theory describing the simplest universality class of critical behavior, 
namely that of the Ising model in a magnetic field \cite{Taniguchi}. The 
quantitative study of the decay processes following the breaking of 
integrability in Ising field theory is the subject of this paper.

The scaling Ising model in two dimensions is described by the action
\EQ
{\cal A}={\cal A}_{0} - \tau\int d^2x\,\var(x) - h \int d^2x\,\sigma(x)\,,
\label{A}
\EN
in which the critical point action ${\cal A}_0$ is perturbed by the energy 
operator $\varepsilon(x)$, with scaling dimension $X_\varepsilon=1$, and by
the spin operator $\sigma(x)$, with scaling dimension $X_\sigma=1/8$. The 
coupling constants\footnote{The mass scale $M$ is the inverse of the 
correlation length.}
\bea
&& \tau\sim M^{2-X_\var}=M\,,\nonumber \\
&& h\sim M^{2-X_\sigma}=M^{15/8}\nonumber
\eea
measure the deviation from critical temperature and the magnetic field, 
respectively, and enter the dimensionless parameter 
\EQ
\eta=\frac{\tau}{|h|^{8/15}}
\label{eta}
\EN
which can be used to label the renormalization group trajectories flowing out 
of the critical point. 

It is well known that the action (\ref{A}) with $h=0$ describes free 
excitations with fermionic statistics. These are ordinary particles in the 
high-temperature phase $\tau>0$ ($\eta=+\infty$), and topological excitations
(kinks) interpolating between the two degenerate vacua of the $\tau<0$ phase 
($\eta=-\infty$) in which the spin reversal simmetry is spontaneously
broken. The evolution of the particle spectrum at $h\neq 0$ was first 
discussed in \cite{McW}. In the limit $\eta\rightarrow -\infty$ the 
particle spectrum exhibits infinitely many discrete levels generated by 
the confinement of the kink-antikink pairs induced by an arbitrarily small 
magnetic field removing the degeneracy among the two vacua of the 
low-temperature phase. Hence, the spectrum in this limit is reproduced by the 
non-relativistic result for a linear confining potential\footnote{The 
spectrum (\ref{nonrel}) was originally obtained in \cite{McW} through the 
study of the analytic structure in momentum space of the spin-spin correlation 
function for small magnetic field.}
\EQ
m_n\simeq 2m+\frac{(2vh)^{2/3}z_n}{m^{1/3}}\,,
\label{nonrel}
\EN
where $m$ is the mass of the kinks, $v$ the spontaneous magnetization and 
$z_n$, $n=1,2,\ldots$, are positive numbers determined by the zeroes of the 
Airy function, Ai$(-z_n)=0$. Of course, the particles with masses $m_n$ larger 
than twice the lightest mass $m_1$ are unstable. It was argued in \cite{McW}
that the number of stable particles descreases as $\eta$ is increased, until
a single asymptotic particle is left in the spectrum at $\eta=+\infty$. This
general pattern has been confirmed by numerical investigation of the spectrum
of the field theory (\ref{A}) \cite{nonint,FZ1}. In particular, it has been 
found that the second and third lightest particles become unstable at 
\EQ
\eta_2=0.333(7)\,,\hspace{1cm}\eta_3\simeq 0.022\,, 
\label{eta23}
\EN
respectively \cite{FZ1}. 
These values of $\eta$, as well as the other normalization-dependent numbers 
we quote in the following, refer to the normalization of the operators in 
which 
\bea
&& \langle\sigma(x)\sigma(0)\rangle\to|x|^{-1/4}\nonumber\\
&& \langle\varepsilon(x)\varepsilon(0)\rangle\to|x|^{-2}
\label{norm}
\eea 
as $|x|\rightarrow 0$. Corrections to the mass $m_1$ for $\eta\rightarrow 
+\infty$ as well as to the spectrum (\ref{nonrel}) have also been computed 
\cite{FZ1,FZ2}.

A.~Zamolodchikov discovered at the end of the eighties that the Ising field
theory (\ref{A}) is integrable for $\tau=0$ and computed the exact 
$S$-matrix through the bootstrap method \cite{Taniguchi}. He found, in 
particular, that the spectrum of the theory on the magnetic axis contains 
eight stable particles $A_a$ ($a=1,\ldots,8$) with masses\footnote{The 
constant relating $m_1$ to the magnetic field was determined in \cite{Fateev}.}
\bea
m_1 &=& (4.40490857..)\,|h|^{8/15}\nonumber \\
m_2 &=& 2 m_1 \cos\frac{\pi}{5} = (1.6180339887..) \,m_1\nonumber\\
m_3 &=& 2 m_1 \cos\frac{\pi}{30} = (1.9890437907..) \,m_1\nonumber\\
m_4 &=& 2 m_2 \cos\frac{7\pi}{30} = (2.4048671724..) \,m_1\nonumber \\
m_5 &=& 2 m_2 \cos\frac{2\pi}{15} = (2.9562952015..) \,m_1\nonumber\\
m_6 &=& 2 m_2 \cos\frac{\pi}{30} = (3.2183404585..) \,m_1\nonumber\\
m_7 &=& 4 m_2 \cos\frac{\pi}{5}\cos\frac{7\pi}{30} = (3.8911568233..) \,m_1\
\nonumber\\
m_8 &=& 4 m_2 \cos\frac{\pi}{5}\cos\frac{2\pi}{15} = (4.7833861168..)\,m_1\,\,.
\nonumber
\eea
Notice that the last five masses all lie above the lowest two-particle 
threshold $2m_1$. Since the Ising model in a magnetic field possesses no 
internal symmetries (so that all particles are completely neutral), the decay
processes $A_a\rightarrow A_1A_1$, $a=4,\ldots,8$, are allowed both by
kinematics and by symmetry. The stability of the particles above threshold, 
however, does not violate any requirement of the $S$-matrix theory as long as 
integrability is present. 

To see this, consider for example the scattering amplitudes of
$A_1$ with itself and with $A_2$. They read\footnote{The energy and momentum of
the particles are parameterized in terms of rapidities as $(p^0,p^1)=
(m_a\cosh\theta,m_a\sinh\theta)$. Relativistic invariant quantities as the
scattering amplitudes depend on rapidity differences only.} \cite{Taniguchi}
\bea
&& S_{11}(\theta)=t_{2/3}(\theta)t_{2/5}(\theta)t_{1/15}(\theta)
\label{S11}\\
&& S_{12}(\theta)=t_{4/5}(\theta)t_{3/5}(\theta)t_{7/15}(\theta)
t_{4/15}(\theta)\,,
\label{S12}
\eea
with 
\EQ
t_\alpha(\theta)=\frac{\tanh\frac{1}{2}(\theta+i\pi\alpha)}
                        {\tanh\frac{1}{2}(\theta-i\pi\alpha)}\,\,.
\label{talpha}
\EN
The simple poles at $\theta=iu_{ab}^c$ in the amplitude $S_{ab}(\theta)$ 
indicate that the particle $A_c$ with mass square
\EQ
m_c^2=m_a^2+m_b^2+2m_am_b\cos u_{ab}^c
\label{mc}
\EN
appears as a bound state in the scattering channel $A_aA_b$. Hence we see that
the scattering channel $A_1A_1$ produces the first three particles as bound 
states, while the channel $A_1A_2$ produces the first four. Figures 1.a and 
1.b show the corresponding analytic structure for the two amplitudes in the 
complex plane of the Mandelstam variable $s$ (square of the center of mass 
energy). As required by the $S$-matrix theory, all the poles corresponding to 
stable bound states lie on the real axis below the lowest unitarity threshold
in the given scattering channel. This is true for all the amplitudes 
$S_{ab}$, $a,b=1,\ldots,8$, of the Ising field theory at $\tau=0$ (see e.g. 
\cite{review} for the full list of amplitudes). When integrability is broken
(i.e. as soon as we move away from $\tau=0$), however, the inelastic channels
and the associated unitarity cuts open up. In particular, the process
$A_1A_2\rightarrow A_1A_1$ acquires a non-zero amplitude, so that the 
threshold located at $s=4m_1^2$ becomes the lowest one also in the $A_1A_2$ 
scattering channel. Since the pole associated to $A_4$ is located above this
threshold, it can no longer remain on the real axis, which in that region is 
now occupied by the new cut. The position of the pole then develops an 
imaginary part which, according to the general requirements for unstable 
particles \cite{ELOP}, is negative and brings the pole through the cut onto 
the unphysical region of the Riemann surface (Fig.~1.c). 

\begin{figure}
\centerline{
\includegraphics[width=9cm]{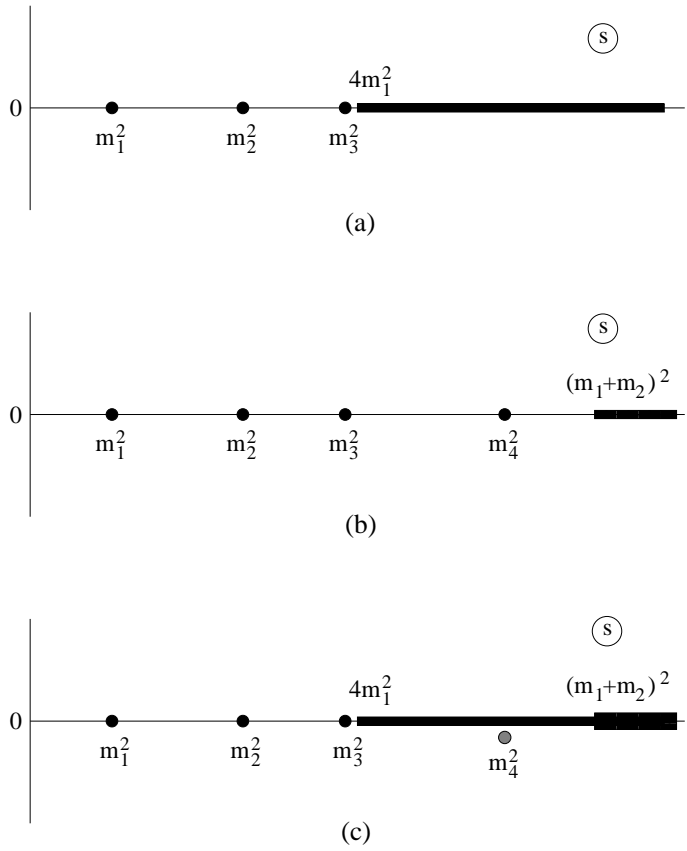}}
\caption{Poles and unitarity cuts for the scattering amplitudes
$S_{11}$ and $S_{12}$ in the integrable case $\tau=0$, (a) and (b), 
respectively, and for $\tau$ slightly different from zero (c). In (c) the 
particle $A_4$ became unstable and the associated pole moved through the cut 
into the unphysical region.}
\end{figure}

\section{Mass corrections and decay widths} 
The change in the position of the poles when the temperature is moved away
from its critical value can be computed perturbatively in $\tau$ within the 
framework of Form Factor Perturbation Theory around integrable models
\cite{nonint}. The form factors of an operator $\Phi(x)$ are its matrix 
elements between the vacuum and the multiparticle asymptotic states. We 
denote them as
\EQ
F^\Phi_{a_1\ldots a_n}(\theta_1,\ldots,\theta_n)=\langle0|\Phi(0)|
A_{a_1}(\theta_1)\ldots A_{a_n}(\theta_n)\rangle\,\,.
\label{formfactors}
\EN
Then the leading corrections to the real and imaginary parts of the particle
masses can be written as (see Fig.~2)
\begin{figure}
\centerline{
\includegraphics[width=8cm]{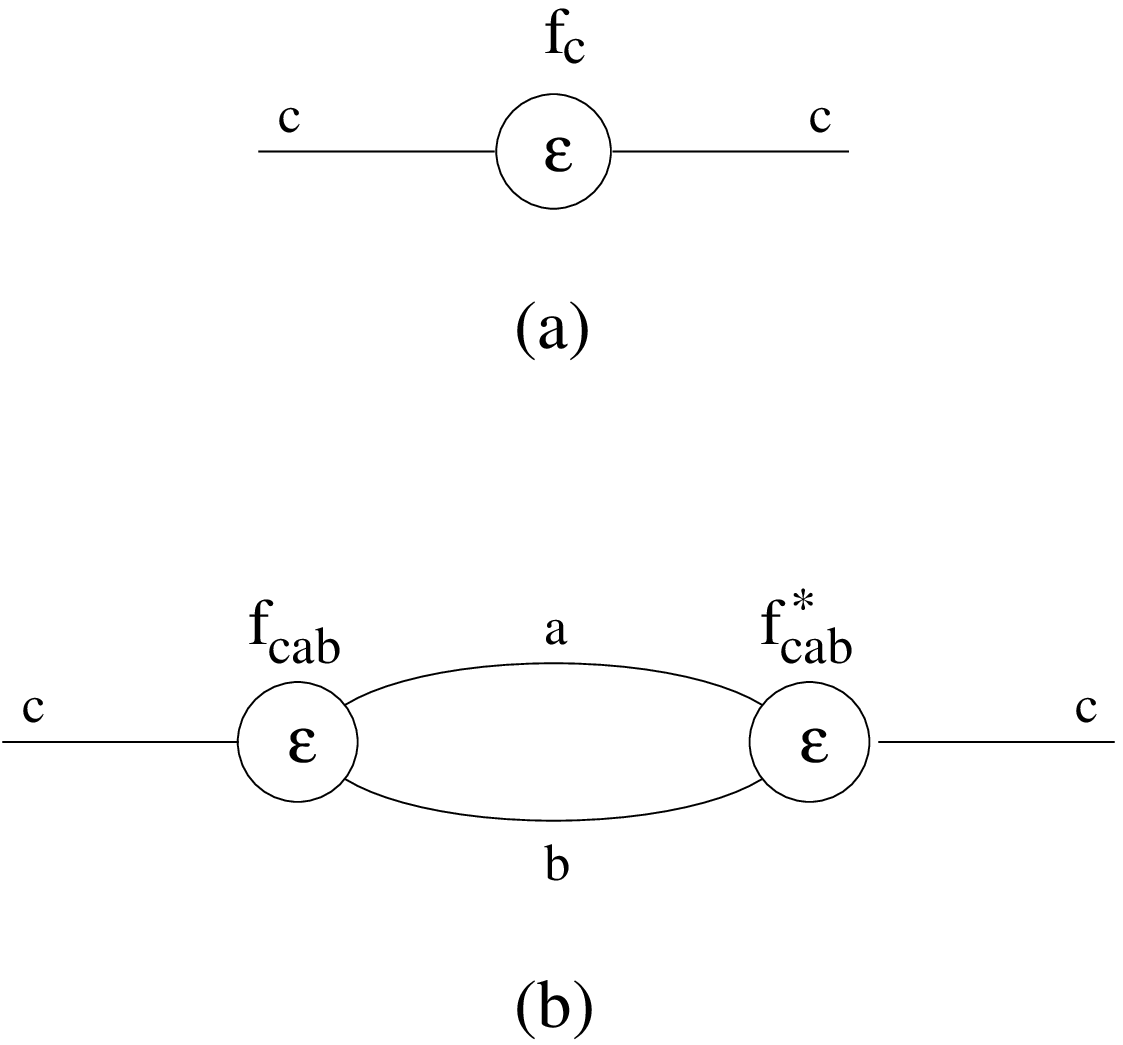}}
\caption{Diagrams determining the leading corrections to the real (a) and
imaginary (b) parts of the masses at small $\tau$. In (b) also the intermediate
particles are on shell and energy-momentum is conserved at each vertex. For 
$c>5$ also diagrams with more than two particles in the intermediate state
contribute to the imaginary part.}
\end{figure}
\EQ
\delta\,\mbox{Re}\,m^2_c\simeq 2\tau\,f_{c}\,,
\label{deltam}
\EN
\EQ
\mbox{Im}\,m_c^2=-\sum_{a\leq b\,,\,m_a+m_b\leq m_c}m_c\,
\Gamma_{c\to ab}\simeq -\tau^2\sum_{a\leq b\,,\,m_a+m_b\leq m_c}
2^{1-\delta_{ab}}\frac{|f_{cab}|^2}
{m_cm_a\left|\sinh\theta^{(cab)}_a\right|}\,,
\label{im}
\EN
where $\Gamma_{c \to ab}$ is the decay width of the particle $A_c$ into the 
channel $A_ a \,A_b$ and 
\EQ
f_c=\left.F^\varepsilon_{cc}(i\pi,0)\right|_{\tau=0}
\EN
\EQ
f_{cab}=\left.F^\varepsilon_{cab}(i\pi,\theta^{(cab)}_a,\theta^{(cab)}_b)
\right|_{\tau=0}\,. 
\EN
The rapidities $\theta^{(cab)}_a$ and $\theta^{(cab)}_b$ are determined by 
energy-momentum conservation for the decay $A_c\rightarrow A_aA_b$ in the 
rest frame of $A_c$. All the masses appearing in the r.h.s. of (\ref{im}) are 
taken at $\tau=0$. For $c>5$ the sum in (\ref{im}) must be completed including
the contributions of the decay channels with more than two particles in the 
final state. Once the decay widths $\Gamma_{c\to ab}$ are known, one can 
determine the lifetime $t_c$ of the unstable particle $A_c$, given by 
\EQ
t_c=\frac{1}{\Gamma_c}\,,\hspace{1cm}
\Gamma_c=\sum_{a\leq b} \Gamma_{c\to ab}\,\,.
\EN
The form factor problem for the Ising field theory in a magnetic field at 
$\tau=0$ has been studied in \cite{immf,DS}, where one and two-particle
form factors have been computed, and these results are reviewed in \cite{review}.
We have now extended the form factor bootstrap to the three-particle case in
order to compute the decay widths (\ref{im}). In the next section we recall the 
main steps of this approach and refer the reader to \cite{review} for details.

\section{Form Factors}
Form factors in an integrable quantum field theory satisfy a set of 
equations \cite{KW,Smirnov} which in the case of neutral particles read
\bea
&& F^\Phi_{a_1\ldots a_ia_{i+1}\ldots a_n}(\theta_1,\ldots,\theta_i,
\theta_{i+1},\ldots,\theta_n)=\nonumber\\
&& \hspace{3.5cm}S_{a_ia_{i+1}}(\theta_i-\theta_{i+1})
F^\Phi_{a_1\ldots a_{i+1}a_{i}\ldots a_n}(\theta_1,\ldots,\theta_{i+1},
\theta_{i},\ldots,\theta_n)
\label{ff1}\\
&& F^\Phi_{a_1\ldots a_n}(\theta_1+2i\pi,\theta_2,\ldots,\theta_n)=
F^\Phi_{a_2\ldots a_na_1}(\theta_2,\ldots,\theta_n,\theta_1)
\label{ff3}\\
&& \mbox{Res}_{\theta_a-\theta_b=iu_{ab}^c}\,
F^\Phi_{aba_1\ldots a_n}(\theta_a,\theta_b,\theta_1,\ldots,\theta_n)=
ig_{ab}^c\,F^\Phi_{ca_1\ldots a_n}(\theta_c,\theta_1,\ldots,\theta_n)
\label{ff2}\\
&& \mbox{Res}_{\theta'=\theta+i\pi}\,
F^\Phi_{aba_1\ldots a_n}(\theta',\theta,\theta_1,\ldots,\theta_n)=
\nonumber\\
&& \hspace{4cm}i\delta_{ab}\left(1-\prod_{j=1}^nS_{a_ja}(\theta_j-\theta)
\right)F^\Phi_{a_1\ldots a_n}(\theta_1,\ldots,\theta_n)\,\,,
\label{ff4}
\eea
with the three-particle couplings $g_{ab}^c$ entering (\ref{ff2}) 
determined from the scattering amplitudes through the relations
\EQ
S_{ab}(\theta\simeq iu_{ab}^c)\simeq
\frac{i(g_{ab}^c)^2}{\theta-iu_{ab}^c}\,\,.
\label{pole}
\EN
A parameterization of the form factors which solves (\ref{ff1}) and {\ref{ff3})
and contains the poles prescribed by (\ref{ff2}) and (\ref{ff4}) is given by
\EQ
F^\Phi_{a_1\ldots a_n}(\theta_1,\ldots,\theta_n)=
Q^\Phi_{a_1\ldots a_n}(\theta_1,\ldots,\theta_n)\prod_{i<j}\frac{
F^{min}_{a_ia_j}(\theta_i-\theta_j)}{\left(e^{\theta_i}+e^{\theta_j}\right)^
{\delta_{a_ia_j}}D_{a_ia_j}(\theta_i-\theta_j)}\,\,.
\label{fn}
\EN
Here 
\EQ
F^{min}_{ab}(\theta)=\left(-i\sinh\frac{\theta}{2}\right)^{\delta_{ab}}
\prod_{\gamma\in{\cal G}_{ab}}\left(T_{\gamma/30}(\theta)
\right)^{p_\gamma}\,\,,
\lab{fmin}
\EN
with
\EQ
T_{\alpha}(\theta)=\exp\left\{2\int_0^\infty\frac{dt}{t}\frac{\cosh\left(
\alpha - \frac{1}{2}\right)t}{\cosh\frac{t}{2}\sinh
t}\sin^2\frac{(i\pi-\theta)t}{2\pi}\right\}
\lab{Talpha}
\EN
satisfying
\EQ
T_\alpha(\theta)=-t_\alpha(\theta)T_\alpha(-\theta)
\EN
\EQ
T_\alpha(\theta+2i\pi)=T_\alpha(-\theta)\,\,
\EN
\EQ
T_\alpha(\theta)\sim\exp{|\theta|/2}\,\,,\hspace{1cm}|\theta|\rightarrow\infty
\,\,.
\EN
The quantities $\gamma$ and $p_\gamma$ in (\ref{fmin}) coincide with those 
entering the expression
\EQ
S_{ab}(\theta)=\prod_{\gamma\in{\cal G}_{ab}}\left(t_{\gamma/30}(\theta)
\right)^{p_\gamma}
\EN
for the amplitudes ($S_{ab}(0)=(-1)^{\delta_{ab}}$). The dynamical poles 
are inserted in (\ref{fn}) through the factors
\EQ
D_{ab}(\theta)=\prod_{\gamma\in{\cal G}_{ab}}\left({\cal P}_{\gamma/30}(\theta)
\right)^{i_\gamma}\left({\cal P}_{1-\gamma/30}(\theta)\right)^{j_\gamma}\,,
\lab{dab}
\EN
where
\EQ
{\cal P}_{\alpha}(\theta)=\frac{\cos\pi\alpha-\cosh\theta}
{2\cos^2\frac{\pi\alpha}{2}}
\lab{polo}
\EN
and
\EQ
\begin{array}{lll}
i_{\gamma} = n+1\,,\hspace{.5cm}& j_{\gamma} = n \,, &
\hspace{.5cm}{\rm if}\hspace{.3cm} p_\gamma=2n+1\\
i_{\gamma} = n \,,\hspace{.5cm}& j_{\gamma} = n \,, &
\hspace{.5cm}{\rm if}\hspace{.3cm} p_\gamma=2n\,\,.
\end{array}
\EN
Multiple poles corresponding to $p_\gamma>1$ are related to multiscattering 
processes \cite{CT} and give rise to residue equations which generalize
(\ref{ff2}) \cite{immf}. 

The operator dependence in (\ref{fn}) is contained in the entire functions
$Q_{a_1\ldots a_n}^\Phi(\theta_1,\ldots,\theta_n)$. For scalar operators 
$\Phi$ with scaling dimension $X_\Phi$, the asymptotic bound \cite{immf}
\EQ
\lim_{|\theta_i|\rightarrow\infty}F^{\Phi}_{a_1\ldots a_n}
(\theta_1,\ldots,\theta_n)\sim\exp(Y_\Phi|\theta_i|)
\label{asymptotics}
\EN
\EQ
Y_\Phi\leq\frac{X_\Phi}{2}\,,
\label{bound}
\EN
together with the invariance under $\theta_j\rightarrow \theta_j+2i\pi$, 
implies that 
the $Q_{a_1\ldots a_n}^\Phi$ are rational functions of the variables 
$x_j\equiv e^{\theta_j}$. In the two-particle case, taking into account 
Lorentz invariance, the
vanishing of the residue (\ref{ff2}) and the identity $F_{a_1a_2}(\theta_1,
\theta_2)=F_{a_2a_1}(\theta_1,\theta_2)$, one can write
\EQ
Q_{a_1a_2}^\Phi(\theta_1,\theta_2)=(x_1+x_2)^{\delta_{a_1a_2}}
P_{a_1a_2}^\Phi(\theta_1,\theta_2)
\EN
with
\EQ
P_{a_1a_2}^\Phi(\theta_1,\theta_2)=\sum_{k=0}^{N_{a_1a_2}} 
c^{\Phi,k}_{a_1a_2}\,\cosh^k(\theta_1-\theta_2)\,\,.
\label{q2}
\EN
In the three-particle case we use the notation
\begin{eqnarray}
{\mathcal Q}^{\Phi}_{a_1 a_2 a_3}(\theta_1,\theta_2,\theta_3) = 
\sum_{\alpha_1,\alpha_2,\alpha_3=\alpha_{1}',\alpha_{2}',
\alpha_{3}'}^{\alpha_{1}'',\alpha_{2}'',\alpha_{3}''}{d}^{\Phi,
\{ \alpha_1, \alpha_2 ,\alpha_3\}}_{ a_1 a_2  a_3}\,x_1^{\alpha_1}
x_2^{\alpha_2} x_3^{\alpha_3}\,\,.
\label{q3}
\end{eqnarray}
The coefficients in (\ref{q2}) and (\ref{q3}) are real and restricted in 
number by the bound (\ref{bound}). In particular, defining 
$\textrm{Deg}_{a_i a_j}$ through
\begin{eqnarray}
\frac{F^{min}_{a_i a_j}(\theta_i-\theta_j)}{D_{a_i a_j}
(\theta_i-\theta_j)} \ \sim \ e^{ \mp \textrm{\small Deg}_{a_i a_j}\theta_i}\,,
\hspace{1cm}\theta_i  \to \pm \infty
\end{eqnarray}
and
\begin{eqnarray}
&& \textrm{Deg}_{i,+} = - \textrm{Deg}_{a_i a_j} -\textrm{Deg}_{a_i a_k}-
\delta_{a_i a_j}-\delta_{a_i a_k} \\
&& \textrm{Deg}_{i,-} =\textrm{Deg}_{a_i a_j} + \textrm{Deg}_{a_i a_k}
\end{eqnarray}
($i,j,k$ a permutation of $1,2,3$), we conclude
\begin{eqnarray}
\alpha_{i}' & = & -\textrm{Deg}_{i,-}\\
\alpha_{i}'' & = & -\textrm{Deg}_{i,+}
\end{eqnarray}
for both $\Phi=\sigma,\varepsilon$ in (\ref{q3}). 
Further constraints on (\ref{q3}) come from symmetry under permutations of 
identical particles
\begin{eqnarray}
{d}^{\Phi,\{\alpha_1,\alpha_2,\alpha_3\}}_{a_i a_i a_j}= 
{d}^{\Phi,\{\alpha_2,\alpha_1,\alpha_3\}}_{a_i a_i a_j}\,,
\end{eqnarray}  
and from Lorentz invariance, which for scalar operators implies
\begin{eqnarray}
\sum_{k=1}^3 \alpha_k = \sum_{1\leq i<j\leq 3} \delta_{a_i a_j}\,\,.
\end{eqnarray}

The form factor equations together with the asymptotic bound allow to show 
that the space of non-trivial relevant scalar operators is two-dimensional
\cite{review}, as expected for the Ising model, but cannot resolve the linear
combination between $\sigma$ and $\varepsilon$. This is achieved imposing the
asymptotic factorization condition
\EQ
\lim_{\alpha\goto+\infty}F^{{\Phi}}_{a_1\ldots a_rb_1\ldots b_{l}}
(\theta_1+\alpha,\ldots,\theta_r+\alpha,\theta_1',\ldots,\theta_l')=
\frac{1}{\langle\Phi\rangle}
F^{{\Phi}}_{a_1\ldots a_{r}}(\theta_1,\ldots,\theta_r)
F^{{\Phi}}_{b_{1}\ldots b_{l}}(\theta_1',\ldots,\theta_l')\,,
\label{cluster}
\EN
which holds, in particular, for relevant scalar scaling operators in theories 
without internal
symmetries \cite{DSC}. Of the two form factor solutions obtained in this way,
that corresponding to $\sigma$ is picked up exploiting the proportionality
between this operator and the trace of the energy-momentum tensor at $\tau=0$
\cite{immf,DS}; the remaining solution must then correspond to $\varepsilon$.
Once the initial conditions of the form factor bootstrap for 
$\Phi=\sigma,\varepsilon$ have been fixed in this way, the functions 
$Q_{a_1\ldots a_n}^\Phi$ in (\ref{fn}) for these operators can be uniquely 
determined using the residue equations. We give in \cite{web} a list of 
results for $n=2$ (which extends that given in \cite{immf,DS}) and for 
$n=3,4$. 

\section{Results}
From the above results we extract the following values to be used in 
(\ref{deltam}) and (\ref{im})
\bea
&& f_{1}=(-17.8933..)\langle\varepsilon\rangle_{\tau=0}
\label{f1}\\
&& f_{2}=(-24.9467..)\langle\varepsilon\rangle_{\tau=0}\\
&& f_{3}=(-53.6799..)\langle\varepsilon\rangle_{\tau=0}\\
&& f_{4}=(-49.3206..)\langle\varepsilon\rangle_{\tau=0}
\label{f4}\\
&& |f_{411}|=(36.73044..)\left|\langle\varepsilon\rangle\right|_{\tau=0}
\label{f411}\\
&& |f_{511}|=(19.16275..)\left|\langle\varepsilon\rangle\right|_{\tau=0}\\
&& |f_{512}|=(11.2183..)\left|\langle\varepsilon\rangle\right|_{\tau=0}
\,\,.
\label{f512}
\eea
Within the normalization (\ref{norm}) the vacuum expectation value appearing 
in these results takes the value \cite{vevs}
\EQ
\langle\varepsilon\rangle_{\tau=0}=(2.00314..)\,|h|^{8/15}\,\,.
\EN
The variation of the mass ratios
\EQ
r_a\,=\,\frac{\mbox{Re}\,m_a}{m_1}\,\,\,,
\EN
close to the magnetic axis, is given by
\EQ
\delta r_a=-\frac{\tau f_1}{m_1 m_a}\left(r_a^2-\frac{f_a}{f_1}\right)+
O(\tau^2)\,\,.
\label{deltar}
\EN
It can be immediately checked that the values (\ref{f1})-(\ref{f4}) imply
positive coefficients for $\tau$ in this equation when $a=1,\ldots,4$, in 
agreement with the expectation that the distance of the stable (unstable) 
particles from the threshold $2m_1$ decreases (increases) as $\eta$ increases.
The mass variations of the first three particles close to the magnetic axis
were measured in \cite{nonint,GR} and fully agree with the predictions of 
the Form Factor Perturbation Theory.

It is tempting to ignore the non-linear corrections to (\ref{deltar}) and
obtain a first-order estimate $\eta_n^{(1)}$ of the values $\eta_n$ at which
the $n$-th particle decays. These are determined by the condition
\EQ
r_n(\eta_n)=2\,,\hspace{1cm}n>1\,\,, 
\EN
and, in this way, one obtains
\bea
&& \eta_2^{(1)} \,\simeq \,0.2\\
&& \eta_3^{(1)}\,\simeq \,0.01\\
&& \eta_4^{(1)}\,\simeq\, -0.2\,\,.
\eea
Comparison with (\ref{eta23}) shows that the first-order estimate captures the
correct order of magnitude of the decay point for $n=2,3$. Notice that, 
if for $n=3$ the mass ratio changes of only $0.5\%$ from $\eta=0$ to $\eta_3$, the
variation of $m_3$ and $m_1$ with respect to $m_1|_{\eta=0}$ in the linear 
approximation is considerably larger. The trajectories corresponding to 
$\eta_2$, $\eta_3$ and to the first-order estimate for $\eta_4$ are shown in 
Fig.~3.

\begin{figure}
\centerline{
\includegraphics[width=10cm]{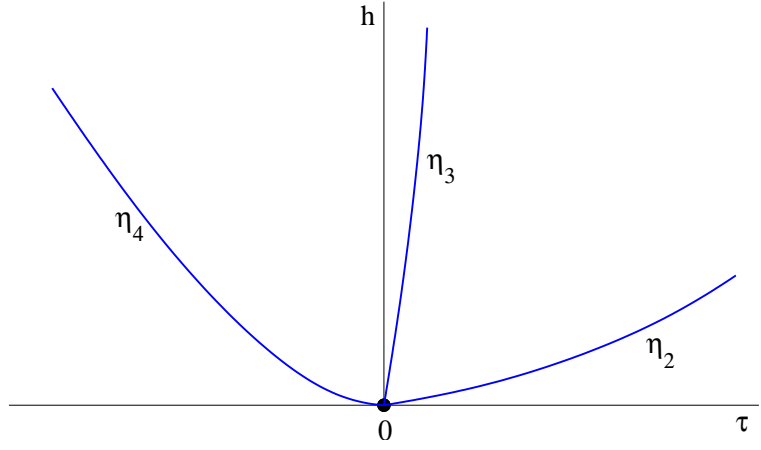}}
\caption{The first few trajectories which divide the $h$-$\tau$ plane
into regions with a different number of stable particles. There are $n$ stable
particles in between the trajectories labeled by $\eta_n$ and $\eta_{n+1}$.
These trajectories densely fill the plane when the negative orizontal axis is
approached ($\eta\to-\infty$).}
\end{figure}

For the imaginary parts of the first two particles above threshold we obtain
\bea
&& \mbox{Im}\,m_4^2\simeq(-840.172..)\left(\frac{\tau
\langle\varepsilon\rangle_{\tau=0}}{m_1}\right)^2=(-173.747..)\,\tau^2\\
&& \mbox{Im}\,m_5^2\simeq(-240.918..)\left(\frac{\tau
\langle\varepsilon\rangle_{\tau=0}}{m_1}\right)^2=(-49.8217..)\,\tau^2\,\,.
\eea
The ratio of lifetimes
\EQ
\lim_{\tau\to 0}\frac{t_4}{t_5}\,=\,\lim_{\tau\to 0}
\frac{m_4\,\mbox{Im}\,m_5^2}{m_5\,\mbox{Im}\,m_4^2}=0.23326..
\label{liferatio}
\EN
is universal. While $A_4$ can only decay into $A_1A_1$, $A_5$ has also the 
$A_1A_2$ channel available. The relevant branching ratios
\EQ
b_{c\to ab}\,=\,\frac{m_c|_{\tau=0}\,\Gamma_{c\to ab}}{|\mbox{Im}\,m_c^2|}
\EN
are 
\EQ
\lim_{\tau\to 0}b_{5\to 11}\,=\,0.47364..\,,\hspace{1cm}
\lim_{\tau\to 0}b_{5\to 12}\,=\,0.52635..\,\,.
\EN

It appears from (\ref{liferatio}) that the lifetime of $A_5$ is more than
four times longer than that of $A_4$, and this seems to contradict, somehow, 
the expectation inherited from accelerator physics that the lifetime of an
unstable particle decreases as its mass increases. Notice, however, that the 
$d$-dimensional phase space for the decay $A_c\to A_aA_b$ is
\EQ
\int\frac{d^{d-1}\vec{p}_a}{p^0_a}\,\,\frac{d^{d-1}\vec{p}_b}{p^0_b}\,\,
\delta^d(p_a+p_b-p_c)\sim\frac{p^{d-3}}{m_c}\,,
\EN
where $p=|\vec{p}_a|=|\vec{p}_b|$ is taken in the center of mass frame. 
For fixed decay products, $p$ increases with $m_c$ and, in $d=2$, it joins the 
factor $m_c$ in the denominator, leading therefore to a suppression of the 
phase space. The results (\ref{f411})--(\ref{f512}) show that such a suppression 
of the decay width is further enhanced by the dynamics in a way that is not 
compensated by the opening of additional decay channels.

In \cite{web} we also give the four-particle form factor 
$F_{1111}^\varepsilon|_{\tau=0}$
which determines the first-order variation to the elastic scattering amplitude
$S_{11}$ through the formula \cite{nonint}
\EQ
\delta S_{11}(\theta_1-\theta_2)\simeq -\frac{i\tau}{m_1^2\sinh(\theta_1-
\theta_2)}\,\lim_{\delta\to 0}F_{1111}^\varepsilon(\theta_1+i\pi+\delta,
\theta_2+i\pi+\delta,\theta_1,\theta_2)\,\,.
\EN
Here the limit is needed to deal with the singularities prescribed by 
(\ref{ff4}).

\section{Numerical studies}

Checks of the analytic results for the particle spectrum may come, in 
particular, from the numerical diagonalization of the Hamiltonian on a 
cylinder geometry, the results for the plane being recovered as the circumference $R$
of the cylinder goes to infinity. For the stable particles of the Ising model
in a magnetic field this kind of analysis has been performed both in the 
continuum \cite{nonint,FZ1,SZ} and on the lattice \cite{GR,CH1,CGR}, always
confirming the analytic predictions. The numerical tests are more complicated 
for unstable particles. 

Qualitatively, the signature of unstable particles on the cylinder is very 
clear \cite{Luscher}. At the integrable point, when the energy levels are 
plotted as a function of $R$, the line corresponding to a particle 
above threshold crosses infinitely many levels which belong to the continuum 
when $R=\infty$ (Fig.~4). Once integrability is broken, this line 
``disappears'' through a removal of level crossings and a reshaping of the
lines associated to stable excitations (Fig.~5). Quantitatively, though, 
it is not so obvious how the decay width can be measured from such a spectrum. 
In absence of a direct measurement of this quantity, one could try however to 
estimate, using the exact matrix elements computed above, the energy splitting 
resulting from the removal of a level crossing taking place at a sufficiently 
large value $R^*$ of the cylinder circumference. 

The usual formula for the first order energy splitting of two states 
$|1\rangle$ and $|2\rangle$, which are degenerate in the unperturbed system, 
is given by 
\EQ
\Delta E=\sqrt{(V_{11}-V_{22})^2+4|V_{12}|^2}\,,
\label{deltae}
\EN
where $V_{ij}=\langle i|V|j\rangle$ are the matrix elements of the perturbing
Hamiltonian $V$. When we put the theory (\ref{A}) on the cylinder and perturb
around $\tau=0$, we have
\EQ
V=-\tau\int_0^Rdx\,\,\varepsilon(x)\,\,.
\EN
For $R^*$ large enough, the corrections coming from the finite volume dynamics 
can be neglected in the first approximation. Hence, the only $R$ dependence comes 
from the change of normalization of the particle states when passing from the plane 
to the cylinder:
\EQ
\langle A_b(p_b)|A_a(p_a)\rangle=2\pi\delta_{ab} p^0_a\delta(p_a^1-p_b^1)
\hspace{.3cm}\longrightarrow\hspace{.3cm}\delta_{ab}
\delta_{p_a^1p_b^1}p^0_aR\,\,.
\EN
By taking into account the normalization $\langle i|j\rangle=\delta_{ij}$
of the states entering (\ref{deltae}), the matrix elements to be used for the
removal of the crossing between the levels corresponding to $A_4$ and $A_1A_1$ 
are then given by 
\bea
&& V_{11}=-\tau R^*\,\frac{f_4}{m_4 R^*} \,\,\,,\\
&& V_{22}=-\tau R^*\,\frac{f_{1111}}{(\frac{m_4}{2}R^*)^2} \,\,\,,\\
&& V_{12}=-\tau R^*\,\frac{f_{411}}{\frac12 (m_4R^*)^{3/2}}\,\,\,,
\eea
where
\bea
f_{1111}&=&\lim_{\delta\to 0}\left.F^\varepsilon_{1111}(\theta_1^{(411)}+
i\pi+\delta,-\theta_1^{(411)}+i\pi+\delta,-\theta_1^{(411)},
\theta_1^{(411)})\right|_{\tau=0}\nonumber \\
&=&(408.78..)\,\langle\varepsilon\rangle_{\tau=0}\,\,.
\eea
For the purpose of comparison with numerical data it can be useful
to consider the universal ratio $\Delta E/\delta E_{gs}(R^*)$, where
\EQ
\delta E_{gs}(R)\,=\,- R\,\tau\,\langle\varepsilon\rangle_{\tau=0}
\EN
is the first order variation of the ground state energy.

Energy spectra on the cylinder can be obtained directly in the continuum
through the truncated conformal space approach \cite{YZtruncation}, in which
the Hamiltonian is diagonalized numerically on a finite dimensional subspace
of the conformal states relative to the critical point. This is 
the method that we have used to obtain the numerical spectrum. By  
including states up to the 5th level of the Verma module of the conformal 
families of the Ising model, our final Hamiltonian has been truncated up 
to 43 states. Even though this truncation of the Hilbert space has proved to 
be successful to measure mass ratio corrections and the change of the 
ground state energy, it fails however to measure with sufficient precision 
the energy splittings which take place above threshold at large values of $R$. 
This was not totally unexpected since it is well known that the accuracy in the
determination of the hamiltonian levels decreases as $R$ and/or the energy 
increase, due to the truncation effects. The remedy to this problem consists, 
of course, in using a larger Hamiltonian, in order to reach a sufficient level 
of precision in the region of large $R$ where we want to measure the energy 
splittings between the higher mass line and the threshold lines. In this 
respect, it would be interesting to make such a check by using the method 
employed in \cite{FZ1}, where the diagonalization is performed on a larger 
number of states of the free fermionic basis corresponding to $h=0$.

\vspace{1cm}
{\bf Acknowledgments.}~~The work of P.G. is supported by the European 
Commission RTN Network EUCLID (contract HPRN-CT-2002-00325); the other authors
are partially supported by this contract. G.D. thanks the 
APCTP in Pohang, South Korea, for hospitality during the final stages of this 
work. G.M. would like to thank CEA, Service de Physique Theorique at Saclay, 
and LPTHE-Jussieu (Paris) for their warm hospitality and for partial financial 
support.

\newpage

\begin{figure}
\centerline{
\includegraphics[width=15cm]{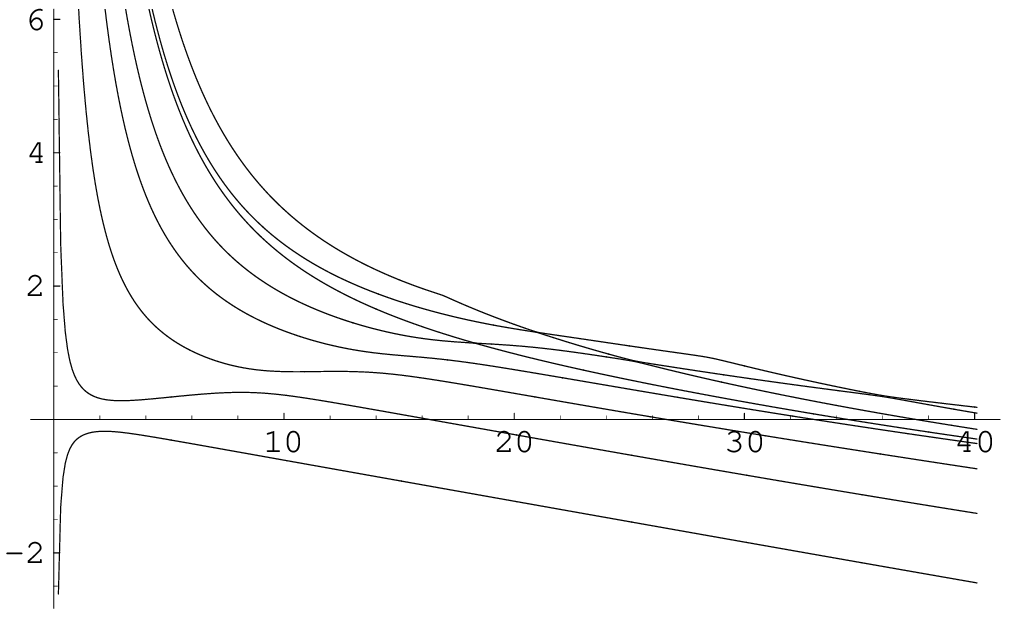}}
\caption{
First eight energy levels of the finite volume Hamiltonian of the scaling Ising
model with magnetic field at critical temperature as functions of the scaling 
variable $r =m_1R$. At $r=40$, starting from the bottom, the levels are 
identified as the ground state, the first three particle states $A_1$, $A_2$ 
and $A_3$, three scattering states $A_1A_1$, the particle above threshold 
$A_4$. Crossings between the line associated to the latter and the scattering 
states are visible around $r=18$, $r=25$ and $r=36$.}
\end{figure}

\begin{figure}
\centerline{
\includegraphics[width=15cm]{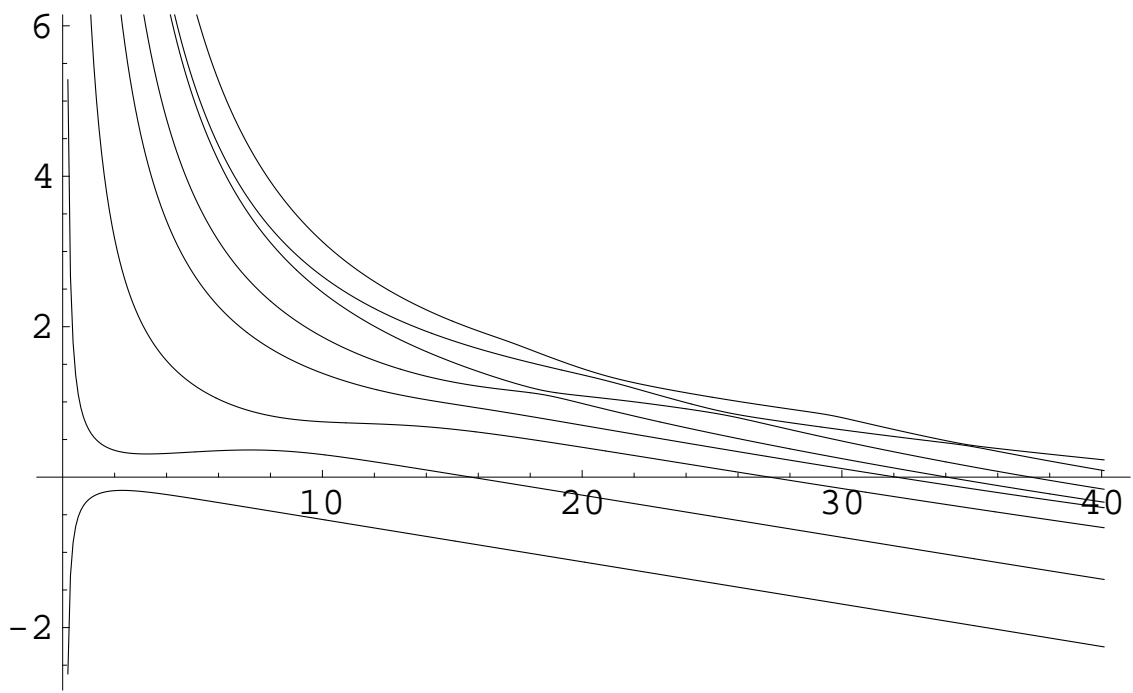}}
\caption{
First eight energy levels of the finite volume Hamiltonian of the scaling 
Ising model with magnetic field slightly away from the critical 
temperature. Observe the splitting of the crossings pointed out in the previous
figure.}
\end{figure}

\end{document}